\documentclass[conference]{IEEEtran}
\IEEEoverridecommandlockouts
\usepackage{cite}
\usepackage{amsmath,amssymb,amsfonts}
\usepackage{amssymb}
\usepackage{algorithmic}
\usepackage{textcomp}
\usepackage[usenames, dvipsnames]{color}
\usepackage[linesnumbered,ruled]{algorithm2e}
\usepackage{subcaption}
\usepackage{graphicx}  
\usepackage {tikz}
\tikzset{>=latex}
\usepackage{xcolor,colortbl}
\definecolor{LayerColor}{RGB}{230, 230, 230}
\definecolor{InOutColor}{RGB}{240, 243, 255}
\definecolor{cellColor}{RGB}{230, 230, 230}

\def\BibTeX{{\rm B\kern-.05em{\sc i\kern-.025em b}\kern-.08em
    T\kern-.1667em\lower.7ex\hbox{E}\kern-.125emX}}
\DeclareMathOperator*{\argmin}{arg\,min}
\DeclareMathOperator*{\argmax}{arg\,max}

\begin{document}
\title{DNN-based Detectors for Massive MIMO Systems with Low-Resolution ADCs}
\author{\IEEEauthorblockN{Ly V. Nguyen\IEEEauthorrefmark{1}, Duy H. N. Nguyen\IEEEauthorrefmark{1}$^,$\IEEEauthorrefmark{2}, and A. Lee Swindlehurst\IEEEauthorrefmark{3}}
\IEEEauthorblockA{\IEEEauthorrefmark{1}Computational Science Research Center, San Diego State University, CA, USA}
\IEEEauthorblockA{\IEEEauthorrefmark{2}Department of Electrical and Computer Engineering,  San Diego State University, CA, USA}
\IEEEauthorblockA{\IEEEauthorrefmark{3}Department of Electrical Engineering and Computer Science, University of California, Irvine, CA, USA.}
Email: vnguyen6@sdsu.edu, duy.nguyen@sdsu.edu, swindle@uci.edu}

\maketitle

\begin{abstract}
Low-resolution analog-to-digital converters (ADCs) have been considered as a practical and promising solution for reducing cost and power consumption in massive Multiple-Input-Multiple-Output (MIMO) systems. Unfortunately, low-resolution ADCs significantly distort the received signals, and thus make data detection much more challenging. In this paper, we develop a new deep neural network (DNN) framework for efficient and low-complexity data detection in low-resolution massive MIMO systems. Based on reformulated maximum likelihood detection problems, we propose two model-driven DNN-based detectors, namely OBMNet and FBMNet, for one-bit and few-bit massive MIMO systems, respectively. The proposed OBMNet and FBMNet detectors have unique and simple structures designed for low-resolution MIMO receivers and thus can be efficiently trained and implemented. Numerical results also show that OBMNet and FBMNet significantly outperform existing detection methods.
\end{abstract}


\section{Introduction}
\label{sec_introduction}
Massive multiple-input multiple-output (MIMO) is considered to be a disruptive technology for 5G-and-beyond networks~\cite{Boccardi2014Five,andrews2014will,Swindlehurst2014Millimeter}. Massive MIMO is capable of boosting the throughput and energy efficiency by several orders of magnitude over conventional MIMO systems~\cite{ngo2013energy,Hoydis2013massive}. A massive MIMO system is equipped with a large number (tens to hundreds) of antennas at the base station and thus requires a large number of radio-frequency (RF) chains, resulting in significant increases in the power consumption and hardware complexity. A practical and promising solution for reducing the power consumption and hardware complexity of such systems is to use low-resolution analog-to-digital converters (ADCs). This is due to the simple structure and low-power consumption of low-resolution ADCs. For example, the simplest architecture involving one-bit ADCs requires only one comparator and does not require an automatic gain control (AGC). Unfortunately, low-resolution ADCs make the system severely nonlinear since the received signals are significantly distorted. The data detection task with low-resolution ADCs therefore becomes even much more challenging compared to conventional full-resolution ADC systems.

There have been a lot of interest and numerous efforts in addressing the data detection problem in massive MIMO systems with low-resolution ADCs, e.g.,~\cite{choi2016near,Mezghani2008Maximum,Jeon2018One,wen2016bayes,nguyen2019linear,Kolomvakis2020Quantized,jeon2019robust,Song2019CRC-Aided,Cho2019OneBitSCSO,Shao2018Iterative}. Maximum-likelihood (ML) detectors for one-bit and few-bit ADCs were derived in~\cite{choi2016near} and~\cite{Mezghani2008Maximum}, respectively. The authors of~\cite{choi2016near} also proposed a so-called near-ML (nML) detection method for large-scale one-bit systems where ML detection is impractical. The ML and nML methods are however non-robust at high signal-to-noise ratios (SNRs) when the channel state information (CSI) is not perfectly known. A one-bit sphere decoding (OSD) technique was proposed in~\cite{Jeon2018One}. However, the OSD technique requires a preprocessing stage whose computational complexity is exponentially proportional to both the number of receive and transmit antennas. The exponential computational complexity of OSD makes it difficult to implement in large-scale MIMO systems. Generalized approximate message passing (GAMP) and Bayes inference were exploited in~\cite{wen2016bayes}, but the resulting method is sophisticated and expensive to implement. Different linear receivers based on the Bussgang decomposition were proposed in~\cite{nguyen2019linear} and~\cite{Kolomvakis2020Quantized} for one-bit and few-bit ADCs, respectively. These linear receivers have lower computational complexity, but often suffer from high detection-error floors, especially with high-dimensional constellations such as $16$-QAM. Several other data detection approaches have also been proposed in~\cite{jeon2019robust,Song2019CRC-Aided,Cho2019OneBitSCSO,Shao2018Iterative}, but they are only applicable in systems where either a cyclic redundancy check (CRC)~\cite{jeon2019robust,Song2019CRC-Aided,Cho2019OneBitSCSO} or an error correcting code such as a low-density parity-check (LDPC) code~\cite{Shao2018Iterative} is available.

Recently, machine learning for MIMO detection has also gained a lot of attention and interest among engineers and researchers. While the deep learning-based detectors in~\cite{Samuel2019Learning,Khani2020Adaptive,Nguyen2020Tabu,Gao2018Sparsely} are designed for MIMO systems with full-resolution ADCs, the learning-based detectors in~\cite{nguyen2019supervised,Jeon2018supervised,Kim2019SemiSupervised} are dedicated to systems with low-resolution ADCs and are ``blind'' in the sense that CSI is not required. However, these blind detection methods are restricted to MIMO systems with a small number of transmit antennas and only low-dimensional constellations. More recently, support vector machines (SVM) were exploited for one-bit MIMO data detection and were shown to achieve better performance than the linear and learning-based receivers ~\cite{Nguyen2020SVM}.

\textit{Contributions:} Motivated by the above discussion, in this paper, we propose novel, efficient, and low-complexity detectors based on deep neural networks (DNNs) for massive MIMO systems with low-resolution ADCs. The proposed DNN-based detectors are also applicable to large-scale systems without the need for CRC or error correcting codes. We first reformulate the ML detection problems by approximating the cumulative distribution function (cdf) of a Gaussian random variable with a Sigmoid activation function, which is a well-known and widely-used activation function in machine learning. Numerical results confirm that the optimal solutions to the reformulated ML detection problems achieve performance that is nearly identical to the original ML detection problems. Based on the reformulated ML detection problems, we then propose model-driven DNN-based detectors, namely OBMNet and FBMNet for one-bit and few-bit massive MIMO systems, respectively. The proposed OBMNet and FBMNet detectors have simple structures that can be implemented in an efficient manner. While each layer of OBMNet has two weight matrices and no bias vector, each layer of FBMNet has two weight matrices and two bias vectors. These weight matrices and bias vectors are adaptive to the channel and the received signal, respectively. In other words, they do not need to be trained and thus result in a much easier training process with much fewer trainable parameters. Once trained, both FMBNet and OBMNet can perform data detection with any new channel realization. Numerical results also show that the proposed OBMNet and FBMNet detectors significantly outperform existing detection methods.

\section{System Model}
\label{sec_system_model}
\begin{figure}[t!]
    \centering
    \includegraphics[width=\linewidth]{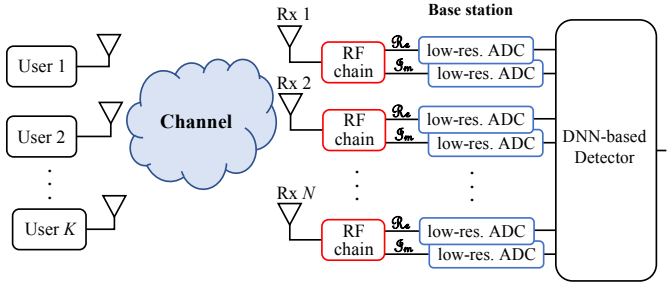}
    \caption{Block diagram of a massive MIMO system with $K$ single-antenna users and an $N$-antenna base station equipped with $2N$ low-resolution ADCs.}
    \label{fig_system_model}
\end{figure}
We consider an uplink massive MIMO system as illustrated in Fig.~\ref{fig_system_model} with $K$ single-antenna users and an $N$-antenna base station, where it is assumed that $N \geq K$. Let $\bar{\mathbf{x}} = [\bar{x}_1, \bar{x}_2, \ldots, \bar{x}_K]^T \in \mathbb{C}^K$ denote the transmitted signal vector, where $\bar{x}_k$ is the signal transmitted from the $k^{\text{th}}$ user under the power constraint $\mathbb{E}[|\bar{x}_k|^2]=1$. The signal $\bar{x}_k$ is drawn from a constellation $\bar{\mathcal{M}}$, e.g, QPSK or $16$-QAM. Let $\bar{\mathbf{H}} \in \mathbb{C}^{N\times K}$ denote the channel, which is assumed to be block flat fading and perfectly known. Let $\bar{\mathbf{r}} = [\bar{r}_1, \bar{r}_2, \ldots, \bar{r}_N]^T \in \mathbb{C}^N$ be the unquantized received signal vector at the base station, which is given by
\begin{equation}
\bar{\mathbf{r}} = \bar{\mathbf{H}}\bar{\mathbf{x}}+\bar{\mathbf{z}},
\label{eq_analog_complex_received_signal}
\end{equation}
where $\bar{\mathbf{z}} = [\bar{z}_1, \bar{z}_2, \ldots,\bar{z}_N]^T \in \mathbb{C}^{N}$ is a noise vector whose elements are assumed to be independent and identically distributed (i.i.d.) as $\mathcal{CN}(0,N_0)$, and $N_0$ is the noise power. Each analog received signal is then quantized by a pair of $b$-bit ADCs. Hence, we have the received signal
\begin{equation}
\bar{\mathbf{y}} = \mathcal{Q}_b(\bar{\mathbf{r}}) = \mathcal{Q}_b\left(\Re\{\bar{\mathbf{r}}\}\right) + j\mathcal{Q}_b\left(\Im\{\bar{\mathbf{r}}\}\right).
\label{eq_quantized_complex_received_signal}
\end{equation}
The operator $\mathcal{Q}_b(\cdot)$ is applied separately to every element of its matrix or vector argument. The SNR is defined as $\rho = 1/N_0$. 

The considered system employs an ADC that performs $b$-bit
uniform scalar quantization. The $b$-bit ADC
model is characterized by a set of $2^b-1$ thresholds denoted
as $\{\tau_1,\ldots,\tau_{2^b-1}\}$. Without loss of generality, we can assume
$-\infty = \tau_0 < \tau_1 <\ldots< \tau_{2^b-1} < \tau_{2^b} = \infty$. Let $\Delta$ be the step size, so the thresholds of the uniform quantizer are given by
\begin{equation}
    \tau_l = (-2^{b-1}+l)\Delta, \; \text{for}\; l \in \mathcal{L}=\{1,\ldots,2^b-1\}.
\end{equation}
The quantization output is defined as
\begin{equation}
    \mathcal{Q}_b(r) = 
    \begin{cases}
    \tau_l - \frac{\Delta}{2} & \text{if}\; r\in(\tau_{l-1},\tau_l]\;\text{with}\;l\in\mathcal{L}\\
    (2^b-1)\frac{\Delta}{2}&\text{if}\;r\in(\tau_{2^b-1},\tau_{2^b}].
    \end{cases}
\end{equation}

\section{Proposed DNN-based Detectors}
\label{sec_DNN-based_detector}
\begin{figure}[t!]
	\centering
	\begin{tikzpicture}
	\node [above] at (0.4,10) {$x^{(0)}_1$};
	\draw [semithick,->] (0,10) to (0.89,10);
	
	\node [above] at (0.4,9) {$x^{(0)}_2$};
	\draw [semithick,->] (0,9) to (0.89,9);
	
	\node [above] at (0.4,7) {$x^{(0)}_{2K}$};
	\draw [semithick,->] (0,7) to (0.89,7);
	
	\draw [fill=LayerColor, thick, rounded corners] (0.9,6.5) rectangle (1.9,10.5);
	\node at (1.4,8.5) {Layer};
	\node at (1.4,8.1) {$1$};
	
	\node [above] at (2.4,10) {$x^{(1)}_1$};
	\draw [semithick,->] (1.9,10) to (2.89,10);
	
	\node [above] at (2.4,9) {$x^{(1)}_2$};
	\draw [semithick,->] (1.9,9) to (2.89,9);
	
	\node [above] at (2.39,7) {$x^{(1)}_{2K}$};
	\draw [semithick,->] (1.9,7) to (2.89,7);
	
	\draw [fill=LayerColor, thick, rounded corners] (2.9,6.5) rectangle (3.9,10.5);
	\node at (3.4,8.5) {Layer};
	\node at (3.4,8.1) {$2$};
	
	\node [above] at (4.4,10) {$x^{(2)}_1$};
	\draw [semithick,->] (3.9,10) to (4.9,10);
	
	\node [above] at (4.4,9) {$x^{(2)}_2$};
	\draw [semithick,->] (3.9,9) to (4.9,9);
	
	\node [above] at (4.39,7) {$x^{(2)}_{2K}$};
	\draw [semithick,->] (3.9,7) to (4.9,7);
	
	\node at (5.3,10) {\textbf{\ldots}};
	\node at (5.3,9) {\textbf{\ldots}};
	\node at (5.3,7) {\textbf{\ldots}};
	
	\node [above] at (6.2,10) {$x^{(L-1)}_1$};
	\draw [semithick,->] (5.6,10) to (6.79,10);
	
	\node [above] at (6.2,9) {$x^{(L-1)}_2$};
	\draw [semithick,->] (5.6,9) to (6.79,9);
	
	\node [above] at (6.15,7) {$x^{(L-1)}_{2K}$};
	\draw [semithick,->] (5.6,7) to (6.79,7);
	
	\draw [fill=LayerColor, thick, rounded corners] (6.8,6.5) rectangle (7.8,10.5);
	\node at (7.3,8.5) {Layer};
	\node at (7.3,8.1) {$L$};
	
	\node [above] at (8.2,10) {$x^{(L)}_1$};
	\draw [semithick,->] (7.8,10) to (8.7,10);
	\node [above] at (8.2,9) {$x^{(L)}_2$};
	\draw [semithick,->] (7.8,9) to (8.7,9);
	\node [above] at (8.2,7) {$x^{(L)}_{2K}$};
	\draw [semithick,->] (7.8,7) to (8.7,7);
	
	\node at (0.4,8.25) {\textbf{\vdots}};
	\node at (2.4,8.25) {\textbf{\vdots}};
	\node at (4.4,8.25) {\textbf{\vdots}};
	\node at (6.2,8.25) {\textbf{\vdots}};
	\node at (8.3,8.25) {\textbf{\vdots}};
	\end{tikzpicture}
	\caption{Overall structure of the proposed OBMNet and FBMNet.}
	\label{fig_overall_network_structure}
\end{figure}
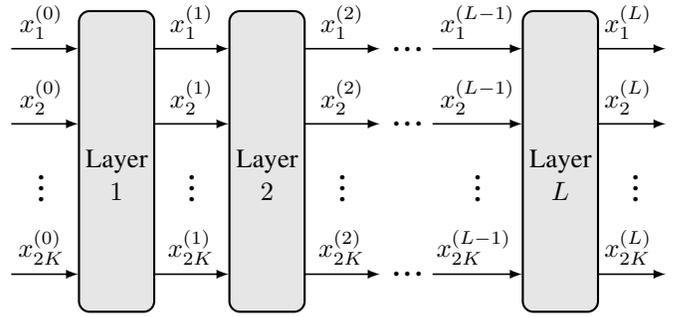

\subsection{FBMNet Detector}
\label{sec_FBMNet_detector}
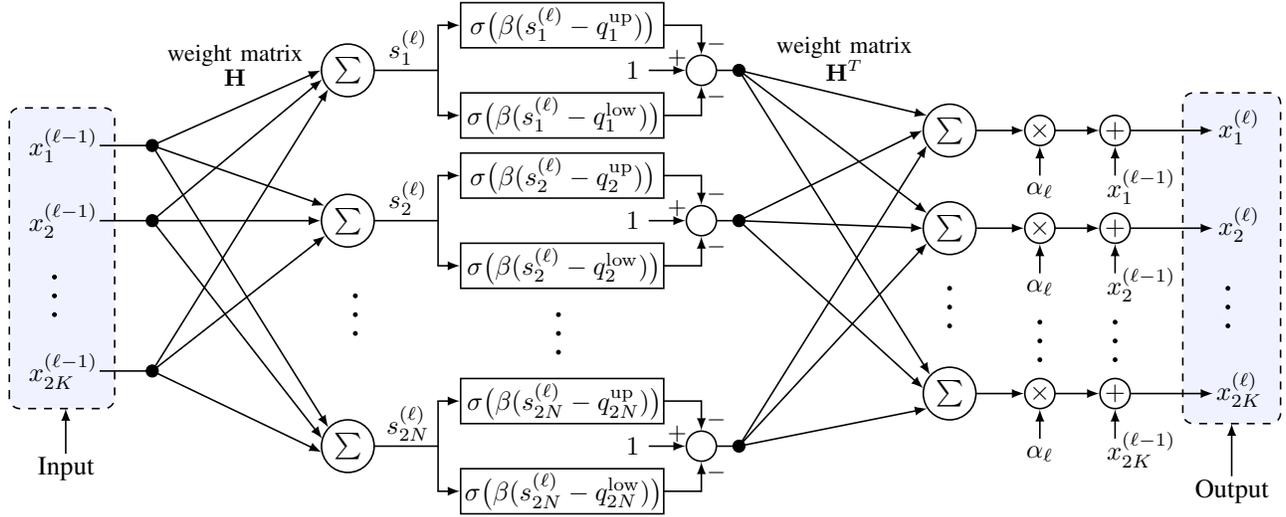
\begin{figure*}[t!]
	\centering
	\begin{tikzpicture}
	\draw [fill=InOutColor, dashed, semithick, rounded corners] (-1.0,6.5) rectangle (0.4,10.5);
	\draw [semithick,->] (-0.25,5.9) to (-0.25,6.5);
	\node at (-0.25,5.7) {Input};
	
	\draw [fill=InOutColor, dashed, semithick, rounded corners] (14.6,6.3) rectangle (15.9,10.7);
	\draw [semithick,->] (15.25,5.6) to (15.25,6.3);
	\node at (15.25,5.4) {Output};
	
	\node at (2,11.2) {\small {weight matrix}};
	\node at (2,10.9) {\small {$\mathbf{H}$}};
	\node at (10.1,11.3) {\small {weight matrix}};
	\node at (10.1,11) {\small {$\mathbf{H}^T$}};
	
	\node [left] at (0.3,10) {$x^{(\ell-1)}_{1}$};
	\draw [semithick] (0.2,10) to (0.9,10);
	\draw [fill] (0.9,10) circle [radius=0.08];
	
	\draw [semithick, ->] (0.9,10) to (3.14,11);
	\draw [semithick, ->] (0.9,9) to (3.16,10.88);
	\draw [semithick, ->] (0.9,7) to (3.24,10.75);
	
	\draw [semithick] (3.5,11) circle [radius=0.35];
	\node at (3.5,11) {$\sum$};
	\draw [semithick] (3.85,11) to (4.7,11);
	\draw [semithick] (4.7,10.4) to (4.7,11.6);
	\draw [semithick,->] (4.69,11.6) to (4.99,11.6);
	\draw [semithick,->] (4.69,10.4) to (4.99,10.4);
    \node at (4.3,11.3) {$s^{(\ell)}_{1}$};
	\draw [semithick] (5,11.3) rectangle (7.7,11.9);
	\node [right] at (5,11.6) {$\sigma\big(\beta(s^{(\ell)}_{1}-q^{\mathrm{up}}_1)\big)$};
	\draw [semithick] (5,10.1) rectangle (7.7,10.7);
	\node [right] at (4.96,10.4) {$\sigma\big(\beta(s^{(\ell)}_{1}-q^{\mathrm{low}}_1)\big)$};
	\draw [semithick] (7.7,11.6) to (8.2,11.6);
	\draw [semithick,->] (8.2,11.61) to (8.2,11.21);
	\draw [semithick] (8.2,11) circle [radius=0.2];
	\draw [semithick] (7.7,10.4) to (8.2,10.4);
	\draw [semithick,->] (8.2,10.39) to (8.2,10.79);
	\draw [semithick,->] (7.5,11) to (7.99,11);
	\node [left] at (7.5,11) {$1$};
	\node at (7.88,11.17) {{\small $+$}};
	\node at (8.4,11.35) {{\small $-$}};
	\node at (8.4,10.65) {{\small $-$}};
	\draw [semithick] (8.4,11) to (8.7,11);
	\draw [fill] (8.7,11) circle [radius=0.08];
	
	\draw [semithick,->] (8.7,11) to (11.2,10.4);
	\draw [semithick,->] (8.7,11) to (11.2,9.1);
	\draw [semithick,->] (8.7,11) to (11.24,6.95);
	
	\draw [semithick] (11.5,10.2) circle [radius=0.35];
	\node at (11.5,10.2) {$\sum$};
	\draw [semithick,->] (11.85,10.2) to (12.49,10.2);
	\draw [semithick] (12.7,10.2) circle [radius=0.2];
	\node at (12.7,10.2) {$\times$};
	\draw [semithick,->] (12.7,9.6) to (12.7,9.99);
	\node at (12.7,9.4) {$\alpha_\ell$};
	\draw [semithick,->] (12.9,10.2) to (13.49,10.2);
	\draw [semithick] (13.7,10.2) circle [radius=0.2];
	\node at (13.7,10.2) {$+$};
	\draw [semithick,->] (13.7,9.6) to (13.7,9.99);
	\node at (14.05,9.45) {$x^{(\ell-1)}_{1}$};
	\draw [semithick,->] (13.9,10.2) to (15,10.2);
	\node at (15.35,10.25) {$x^{(\ell)}_{1}$};
	
	\node [left] at (0.3,9) {$x^{(\ell-1)}_{2}$};
	\draw [semithick] (0.2,9) to (0.9,9);
	\draw [fill] (0.9,9) circle [radius=0.08];
	
	\draw [semithick, ->] (0.9,10) to (3.19,9.2);
	\draw [semithick, ->] (0.9,9) to (3.14,9);
	\draw [semithick, ->] (0.9,7) to (3.2,8.8);
	
	\draw [semithick] (3.5,9) circle [radius=0.35];
	\node at (3.5,9) {$\sum$};
	\draw [semithick] (3.85,9) to (4.7,9);
	\draw [semithick] (4.7,8.4) to (4.7,9.6);
	\draw [semithick,->] (4.69,9.6) to (4.99,9.6);
	\draw [semithick,->] (4.69,8.4) to (4.99,8.4);
    \node at (4.3,9.3) {$s^{(\ell)}_{2}$};
	\draw [semithick] (5,9.3) rectangle (7.7,9.9);
	\node [right] at (5,9.6) {$\sigma\big(\beta(s^{(\ell)}_{2}-q^{\mathrm{up}}_2)\big)$};
	\draw [semithick] (5,8.1) rectangle (7.7,8.7);
	\node [right] at (4.96,8.4) {$\sigma\big(\beta(s^{(\ell)}_{2}-q^{\mathrm{low}}_2)\big)$};
	\draw [semithick] (7.7,9.6) to (8.2,9.6);
	\draw [semithick,->] (8.2,9.61) to (8.2,9.21);
	\draw [semithick] (8.2,9) circle [radius=0.2];
	\draw [semithick] (7.7,8.4) to (8.2,8.4);
	\draw [semithick,->] (8.2,8.39) to (8.2,8.79);
	\draw [semithick,->] (7.5,9) to (7.99,9);
	\node [left] at (7.5,9) {$1$};
	\node at (7.88,9.17) {{\small $+$}};
	\node at (8.4,9.35) {{\small $-$}};
	\node at (8.4,8.65) {{\small $-$}};
	\draw [semithick] (8.4,9) to (8.7,9);
	\draw [fill] (8.7,9) circle [radius=0.08];
	
	\draw [semithick,->] (8.7,9) to (11.14,10.2);
	\draw [semithick,->] (8.7,9) to (11.15,8.9);
	\draw [semithick,->] (8.7,9) to (11.15,6.77);
	
	\draw [semithick] (11.5,8.9) circle [radius=0.35];
	\node at (11.5,8.9) {$\sum$};
	\draw [semithick,->] (11.85,8.9) to (12.49,8.9);
	\draw [semithick] (12.7,8.9) circle [radius=0.2];
	\node at (12.7,8.9) {$\times$};
	\draw [semithick,->] (12.7,8.3) to (12.7,8.69);
	\node at (12.7,8.1) {$\alpha_\ell$};
	\draw [semithick,->] (12.9,8.9) to (13.49,8.9);
	\draw [semithick] (13.7,8.9) circle [radius=0.2];
	\node at (13.7,8.9) {$+$};
	\draw [semithick,->] (13.7,8.3) to (13.7,8.69);
	\node at (14.05,8.15) {$x^{(\ell-1)}_{2}$};
	\draw [semithick,->] (13.9,8.9) to (15,8.9);
	\node at (15.35,8.95) {$x^{(\ell)}_{2}$};
	
	\node [left] at (0.3,7) {$x^{(\ell-1)}_{2K}$};
	\draw [semithick] (0.2,7) to (0.9,7);
	\draw [fill] (0.9,7) circle [radius=0.08];
	
	\draw [semithick, ->] (0.9,10) to (3.24,6.25);
	\draw [semithick, ->] (0.9,9) to (3.16,6.1);
	\draw [semithick, ->] (0.9,7) to (3.15,5.92);
	
	\draw [semithick] (3.5,6) circle [radius=0.35];
	\node at (3.5,6) {$\sum$};
	\draw [semithick] (3.85,6) to (4.7,6);
	\draw [semithick] (4.7,5.4) to (4.7,6.6);
	\draw [semithick,->] (4.69,6.6) to (4.99,6.6);
	\draw [semithick,->] (4.69,5.4) to (4.99,5.4);
    \node at (4.3,6.3) {$s^{(\ell)}_{2N}$};
	\draw [semithick] (5,6.3) rectangle (7.7,6.9);
	\node [right] at (4.96,6.6) {$\sigma\big(\beta(s^{(\ell)}_{2N}-q^{\mathrm{up}}_{2N})\big)$};
	\draw [semithick] (5,5.1) rectangle (7.7,5.7);
	\node [right] at (4.93,5.4) {$\sigma\big(\beta(s^{(\ell)}_{2N}-q^{\mathrm{low}}_{2N})\big)$};
	\draw [semithick] (7.7,6.6) to (8.2,6.6);
	\draw [semithick,->] (8.2,6.61) to (8.2,6.21);
	\draw [semithick] (8.2,6) circle [radius=0.2];
	\draw [semithick] (7.7,5.4) to (8.2,5.4);
	\draw [semithick,->] (8.2,5.39) to (8.2,5.79);
	\draw [semithick,->] (7.5,6) to (7.99,6);
	\node [left] at (7.5,6) {$1$};
	\node at (7.88,6.17) {{\small $+$}};
	\node at (8.4,6.35) {{\small $-$}};
	\node at (8.4,5.65) {{\small $-$}};
	\draw [semithick] (8.4,6) to (8.7,6);
	\draw [fill] (8.7,6) circle [radius=0.08];
	
	\draw [semithick,->] (8.7,6) to (11.2,10);
	\draw [semithick,->] (8.7,6) to (11.2,8.7);
	\draw [semithick,->] (8.7,6) to (11.2,6.5);
	
	\draw [semithick] (11.5,6.7) circle [radius=0.35];
	\node at (11.5,6.7) {$\sum$};
	\draw [semithick,->] (11.85,6.7) to (12.49,6.7);
	\draw [semithick] (12.7,6.7) circle [radius=0.2];
	\node at (12.7,6.7) {$\times$};
	\draw [semithick,->] (12.7,6.1) to (12.7,6.49);
	\node at (12.7,5.9) {$\alpha_\ell$};
	\draw [semithick,->] (12.9,6.7) to (13.49,6.7);
	\draw [semithick] (13.7,6.7) circle [radius=0.2];
	\node at (13.7,6.7) {$+$};
	\draw [semithick,->] (13.7,6.1) to (13.7,6.49);
	\node at (14.05,5.95) {$x^{(\ell-1)}_{2K}$};
	\draw [semithick,->] (13.9,6.7) to (15,6.7);
	\node at (15.35,6.75) {$x^{(\ell)}_{2K}$};
	
	\draw [fill] (-0.4,8.25) circle [radius=0.025];
	\draw [fill] (-0.4,8) circle [radius=0.025];
	\draw [fill] (-0.4,7.75) circle [radius=0.025];
	
	\draw [fill] (3.6,8) circle [radius=0.025];
	\draw [fill] (3.6,7.75) circle [radius=0.025];
	\draw [fill] (3.6,7.5) circle [radius=0.025];
	
	\draw [fill] (6.3,7.75) circle [radius=0.025];
	\draw [fill] (6.3,7.5) circle [radius=0.025];
	\draw [fill] (6.3,7.25) circle [radius=0.025];

	\draw [fill] (11.5,8.1) circle [radius=0.025];
	\draw [fill] (11.5,7.85) circle [radius=0.025];
	\draw [fill] (11.5,7.6) circle [radius=0.025];

	\draw [fill] (12.7,7.65) circle [radius=0.025];
	\draw [fill] (12.7,7.4) circle [radius=0.025];
	\draw [fill] (12.7,7.15) circle [radius=0.025];

	\draw [fill] (13.7,7.65) circle [radius=0.025];
	\draw [fill] (13.7,7.4) circle [radius=0.025];
	\draw [fill] (13.7,7.15) circle [radius=0.025];

	\draw [fill] (15.2,8.1) circle [radius=0.025];
	\draw [fill] (15.2,7.85) circle [radius=0.025];
	\draw [fill] (15.2,7.6) circle [radius=0.025];
	\end{tikzpicture}
	\caption{Specific structure of layer $\ell$ of FBMNet. The weight matrices and the bias vectors are defined by the channel and the received signal, respectively.}
	\label{fig_fewbit_structure_layer_l}
\end{figure*}

In this section, we propose a DNN-based detector referred to as FBMNet for few-bit ($b>1$) massive MIMO systems. The extreme case of $1$-bit ADCs will be considered later, and special DNN-based detector for this case, referred to as OBMNet, will be proposed. For convenience in later derivations, we convert~\eqref{eq_analog_complex_received_signal} and~\eqref{eq_quantized_complex_received_signal} into the real domain as follows:
\begin{equation}
\mathbf{y} = \mathcal{Q}_b\left(\mathbf{H}\mathbf{x} + \mathbf{z}\right),
\label{eq_quantized_real_received_signal}
\end{equation}
where
\begin{align*}
\mathbf{y} &= \begin{bmatrix}
\Re \{\bar{\mathbf{y}}\} \\ \Im \{\bar{\mathbf{y}}\}
\end{bmatrix},\ \mathbf{x} = \begin{bmatrix}
\Re \{\bar{\mathbf{x}}\} \\ \Im \{\bar{\mathbf{x}}\}
\end{bmatrix}, \
\mathbf{z} = \begin{bmatrix}
\Re \{\bar{\mathbf{z}}\} \\ \Im \{\bar{\mathbf{z}}\}
\end{bmatrix}, \ \text{and}\\
\mathbf{H} &= \begin{bmatrix}
\Re \{\bar{\mathbf{H}}\} & -\Im \{\bar{\mathbf{H}}\}\\
\Im \{\bar{\mathbf{H}}\} & \Re \{\bar{\mathbf{H}}\}
\end{bmatrix}.
\end{align*}
Note that $\mathbf{y}\in \mathbb{R}^{2N}$, $\mathbf{x}\in \mathbb{R}^{2K}$, $\mathbf{z}\in \mathbb{R}^{2N}$, and $\mathbf{H}\in \mathbb{R}^{2N\times2K}$. We also denote $\mathbf{y} = [y_1, \ldots, y_{2N}]^T$ and $\mathbf{H} = [\mathbf{h}_1, \ldots, \mathbf{h}_{2N}]^T$. 

Let $t^\mathrm{up}_i=\sqrt{2\rho}(q^{\mathrm{up}}_i-\mathbf{h}_i^T\mathbf{x})$ and $t^\mathrm{low}_i=\sqrt{2\rho}(q^{\mathrm{low}}_i-\mathbf{h}_i^T\mathbf{x})$, where
\begin{align*}
    q^{\mathrm{up}}_i &= 
    \begin{cases}
    y_i+\frac{\Delta}{2} & \text{if}\; y_i<\tau_{2^b-1}\\
    \infty & \text{otherwise},
    \end{cases}\\
    q^{\mathrm{low}}_i &= 
    \begin{cases}
    y_i-\frac{\Delta}{2} & \text{if}\; y_i>\tau_1\\
    -\infty & \text{otherwise}.
    \end{cases}
\end{align*}
Hence, $q^{\mathrm{up}}_i$ and $q^{\mathrm{low}}_i$ are the upper and lower quantization thresholds of the bin to which $y_i$ belongs. The likelihood function of $\mathbf{x}$ given the received signal $\mathbf{y}$ can be written as~\cite{Mezghani2008Maximum}
\begin{equation}
    \mathcal{P}(\mathbf{x}) = \prod_{i=1}^{2N}\Big(\Phi \left(t^\mathrm{up}_i\right)-\Phi \left(t^\mathrm{low}_i\right)\Big)
\end{equation}
where $\Phi(t) = \int_{-\infty}^{t}\frac{1}{\sqrt{2\pi}}e^{-\frac{\tau^2}{2}}d\tau$ is the cdf of the standard Gaussian random variable. 
The ML detection problem based on the log-likelihood function is defined as follows:
\begin{equation}
\hat{\mathbf{x}}_{\mathtt{ML}} = \argmax_{\bar{\mathbf{x}}\in \bar{\mathcal{M}}^{K}}\; \sum_{i=1}^{2N}\log\left[\Phi \left(t^\mathrm{up}_i\right)-\Phi \left(t^\mathrm{low}_i\right)\right].
\label{eq_conventional_logML_detection}
\end{equation}

We note that an optimal solution for ML detection requires an exhaustive search over $\bar{\mathcal{M}}^{K}$. In addition, there is no closed-form expression for $\Phi(\cdot)$, which complicates the evaluation in \eqref{eq_conventional_logML_detection}. Thus, we first exploit a result in~\cite{bowling2009logistic}, which shows that the function $\Phi(t)$ can be accurately approximated by the Sigmoid function $\sigma(t) = 1/(1+e^{-t})$. More specifically, 
\begin{equation}
	\Phi(t) \approx \sigma(ct) = \frac{1}{1+e^{-ct}},
	\label{eq_approximate_Phi_as_Sigma}
\end{equation}
where $c = 1.702$ is a constant. It was shown in~\cite{bowling2009logistic} that $|\Phi(t)-\sigma(ct)|\leq 0.0095$, $\forall t\in \mathbb{R}$. Then an approximate version of the log-likelihood function of $\mathcal{P}(\mathbf{x})$ can be written as follows:
\begin{align}
    \tilde{\mathcal{P}}(\mathbf{x}) &\approx \sum_{i=1}^{2N}\log\left[\frac{1}{1+e^{-ct^\mathrm{up}_i}}-\frac{1}{1+e^{-ct^\mathrm{low}_i}}\right]\\
    &= \sum_{i=1}^{2N}\Big[\log\Big(e^{-ct^\mathrm{low}_i}-e^{-ct^\mathrm{up}_i}\Big) - \notag\\ & \qquad \quad \;\; \log\left(1+e^{-ct^\mathrm{up}_i}\right)- \log\left(1+e^{-ct^\mathrm{low}_i}\right)\Big].
\end{align}
The reformulated ML detection problem is thus
\begin{equation}
\hat{\mathbf{x}}_{\mathtt{ML}} = \argmax_{\bar{\mathbf{x}}\in \bar{\mathcal{M}}^{K}}\; \tilde{\mathcal{P}}(\mathbf{x}).
\label{eq_reformulated_logML_detection}
\end{equation}

\begin{figure*}[t!]
	\centering
	\begin{tikzpicture}
	\draw [fill=InOutColor, dashed, semithick, rounded corners] (-1.4,6.5) rectangle (0.1,10.5);
	\draw [semithick,->] (-0.65,5.9) to (-0.65,6.5);
	\node at (-0.65,5.7) {Input};
	
	\draw [fill=InOutColor, dashed, semithick, rounded corners] (12.3,6.5) rectangle (13.7,10.5);
	\draw [semithick,->] (13,5.9) to (13,6.5);
	\node at (13,5.7) {Output};
	
	\node [left] at (-0.1,10) {$x^{(\ell-1)}_1$};
	\draw [semithick] (-0.1,10) to (0.6,10);
	\draw [fill] (0.6,10) circle [radius=0.08];
	
	\draw [semithick] (3.6,11) circle [radius=0.35];
	\node at (3.6,11) {$\sum$};
	\draw [semithick,->] (3.95,11) to (4.6,11);
	\node at (4.3,11.3) {$s^{(\ell)}_{1}$};
	\draw [semithick] (4.6,10.7) rectangle (6,11.3);
	\node at (5.3,11) {$\sigma\big(\beta s_1^{(\ell)}\big)$};
	
	\draw [semithick] (9,10.1) circle [radius=0.35];
	\node at (9,10.1) {$\sum$};
	\draw [semithick,->] (6,11) to (8.67,10.2);
	\draw [semithick,->] (6,9.5) to (8.65,10.1);
	\draw [semithick,->] (6,6) to (8.7,9.9);
	
	\draw [semithick, ->] (0.6,10) to (3.24,11);
	\draw [semithick, ->] (0.6,9) to (3.26,10.88);
	\draw [semithick, ->] (0.6,7) to (3.34,10.75);
	
	\draw [semithick, ->] (9.35,10.1) to (9.99,10.1);
	\draw [semithick] (10.2,10.1) circle [radius=0.2];
	\node at (10.2,10.1) {\small {$\times$}};
	\draw [semithick,->] (10.2,9.6) to (10.2,9.89);
	\node at (10.2,9.45) {$\alpha_{\ell}$};
	\draw [semithick, ->] (10.4,10.1) to (11.09,10.1);
	\draw [semithick] (11.3,10.1) circle [radius=0.2];
	\draw [semithick, ->] (11.3,9.6) to (11.3,9.89);
	\node at (11.65,9.49) {$x^{(\ell-1)}_1$};
	\node at (11.3,10.1) {\small{$+$}};
	\draw [semithick, ->] (11.5,10.1) to (12.7,10.1);
	\node [right] at (12.7,10.1) {$x^{(\ell)}_1$};
	
	\node [left] at (-0.1,9) {$x^{(\ell-1)}_2$};
	\draw [semithick] (-0.1,9) to (0.6,9);
	\draw [fill] (0.6,9) circle [radius=0.08];
	
	\draw [semithick] (3.6,9.5) circle [radius=0.35];
	\node at (3.6,9.5) {$\sum$};
	\draw [semithick,->] (3.95,9.5) to (4.6,9.5);
	\node at (4.3,9.8) {$s^{(\ell)}_{2}$};
	\draw [semithick] (4.6,9.2) rectangle (6,9.8);
	\node at (5.3,9.5) {$\sigma\big(\beta s_2^{(\ell)}\big)$};
	
	\draw [semithick] (9,9) circle [radius=0.35];
	\node at (9,9) {$\sum$};
	\draw [semithick,->] (6,11) to (8.67,9.1);
	\draw [semithick,->] (6,9.5) to (8.65,9);
	\draw [semithick,->] (6,6) to (8.7,8.8);
	
	\draw [semithick, ->] (0.6,10) to (3.26,9.62);
	\draw [semithick, ->] (0.6,9) to (3.24,9.49);
	\draw [semithick, ->] (0.6,7) to (3.3,9.3);
	
	\draw [semithick, ->] (9.35,9) to (9.99,9);
	\draw [semithick] (10.2,9) circle [radius=0.2];
	\node at (10.2,9) {\small {$\times$}};
	\draw [semithick,->] (10.2,8.5) to (10.2,8.79);
	\node at (10.2,8.35) {$\alpha_{\ell}$};
	\draw [semithick, ->] (10.4,9) to (11.09,9);
	\draw [semithick] (11.3,9) circle [radius=0.2];
	\draw [semithick, ->] (11.3,8.5) to (11.3,8.79);
	\node at (11.65,8.4) {$x^{(\ell-1)}_2$};
	\node at (11.3,9) {\small {$+$}};
	\draw [semithick, ->] (11.5,9) to (12.7,9);
	\node [right] at (12.7,9) {$x^{(\ell)}_2$};
	
	\node [left] at (-0.1,7) {$x^{(\ell-1)}_{2K}$};
	\draw [semithick] (-0.1,7) to (0.6,7);
	\draw [fill] (0.6,7) circle [radius=0.08];
	
	\draw [semithick] (3.6,6) circle [radius=0.35];
	\node at (3.6,6) {$\sum$};
	\draw [semithick,->] (3.95,6) to (4.6,6);
	\node at (4.25,6.35) {$s^{(\ell)}_{2N}$};
	\draw [semithick] (4.6,5.7) rectangle (6,6.3);
	\node at (5.3,6) {$\sigma\big(\beta s_{2N}^{(\ell)}\big)$};
	
	\draw [semithick] (9,7) circle [radius=0.35];
	\node at (9,7) {$\sum$};
	\draw [semithick,->] (6,11) to (8.7,7.2);
	\draw [semithick,->] (6,9.5) to (8.65,7.05);
	\draw [semithick,->] (6,6) to (8.7,6.8);
	
	\draw [semithick, ->] (0.6,10) to (3.37,6.28);
	\draw [semithick, ->] (0.6,9) to (3.26,6.1);
	\draw [semithick, ->] (0.6,7) to (3.25,5.92);
	
	\draw [semithick, ->] (9.35,7) to (9.99,7);
	\draw [semithick] (10.2,7) circle [radius=0.2];
	\node at (10.2,7) {\small {$\times$}};
	\draw [semithick,->] (10.2,6.5) to (10.2,6.79);
	\node at (10.2,6.35) {$\alpha_{\ell}$};
	\draw [semithick, ->] (10.4,7) to (11.09,7);
	\draw [semithick] (11.3,7) circle [radius=0.2];
	\draw [semithick, ->] (11.3,6.5) to (11.3,6.79);
	\node at (11.65,6.4) {$x^{(\ell-1)}_{2K}$};
	\node at (11.3,7) {\small {$+$}};
	\draw [semithick, ->] (11.5,7) to (12.7,7);
	\node [right] at (12.7,7) {$x^{(\ell)}_{2K}$};
	
	\draw [fill] (-0.7,8.25) circle [radius=0.025];
	\draw [fill] (-0.7,8) circle [radius=0.025];
	\draw [fill] (-0.7,7.75) circle [radius=0.025];
	
	\draw [fill] (3.6,8.25) circle [radius=0.025];
	\draw [fill] (3.6,8) circle [radius=0.025];
	\draw [fill] (3.6,7.75) circle [radius=0.025];
	
	\draw [fill] (5.1,8.25) circle [radius=0.025];
	\draw [fill] (5.1,8) circle [radius=0.025];
	\draw [fill] (5.1,7.75) circle [radius=0.025];
	
	\draw [fill] (9,8.25) circle [radius=0.025];
	\draw [fill] (9,8) circle [radius=0.025];
	\draw [fill] (9,7.75) circle [radius=0.025];
	
	\draw [fill] (10.2,8) circle [radius=0.025];
	\draw [fill] (10.2,7.75) circle [radius=0.025];
	\draw [fill] (10.2,7.5) circle [radius=0.025];
	
	\draw [fill] (11.3,8) circle [radius=0.025];
	\draw [fill] (11.3,7.75) circle [radius=0.025];
	\draw [fill] (11.3,7.5) circle [radius=0.025];
	
	\draw [fill] (13,8.25) circle [radius=0.025];
	\draw [fill] (13,8) circle [radius=0.025];
	\draw [fill] (13,7.75) circle [radius=0.025];
	
	\node at (1.8,11.2) {\small {weight matrix}};
	\node at (1.8,10.9) {\small {$-\mathbf{G}$}};
	
	\node at (7.5,11.2) {\small {weight matrix}};
	\node at (7.5,10.9) {\small {$\mathbf{G}^T$}};
	\end{tikzpicture}
	\caption{Specific structure of layer $\ell$ of OBMNet. The weight matrices are defined by the channel and the received signal. There is no bias vector.}
	\label{fig_1bit_structure_layer_l}
\end{figure*}
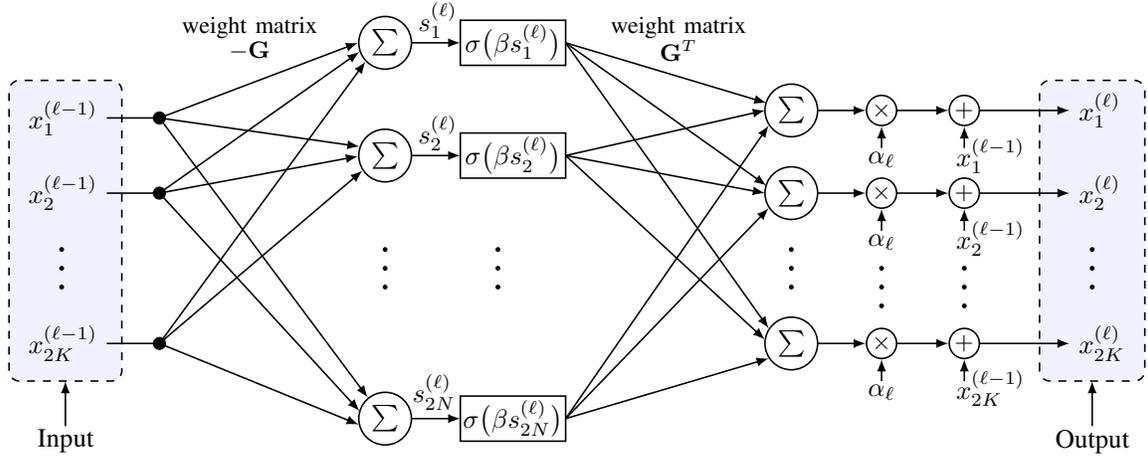

Note that an optimal solution to problem \eqref{eq_reformulated_logML_detection} still requires an exhaustive search over $\bar{\mathcal{M}}^{K}$. Thus, we relax the constraint $\bar{\mathbf{x}}\in \bar{\mathcal{M}}^{K}$ in~\eqref{eq_reformulated_logML_detection} to $\bar{\mathbf{x}}\in \mathbb{C}^{K}$ and solve the following optimization problem:
\begin{equation}
\begin{aligned}
& \underset{\bar{\mathbf{x}}\in \mathbb{C}^{K}}{\operatorname{maximize}}
& & \tilde{\mathcal{P}}(\mathbf{x}).
\end{aligned}
\label{eq_relaxed_logML_detection}
\end{equation}
The gradient of $\tilde{\mathcal{P}}(\mathbf{x})$ is 
\begin{align}
    \nabla\mathcal{P}(\mathbf{x})&=\sum_{i=1}^{2N}c\sqrt{2\rho}\,\mathbf{h}_i\left(1-\frac{1}{1+e^{ct^\mathrm{up}_i}}-\frac{1}{1+e^{ct^\mathrm{low}_i}}\right)\\
    &= c\sqrt{2\rho}\,\mathbf{H}^T\Big[\mathbf{1}-\sigma\left(c\sqrt{2\rho}\left(\mathbf{H}\mathbf{x} - \mathbf{q}^{\mathrm{up}}\right)\right)-\notag\\
    &\qquad\qquad\qquad\qquad\sigma\left(c\sqrt{2\rho}\left(\mathbf{H}\mathbf{x}-\mathbf{q}^{\mathrm{low}}\right)\right)\Big],
\end{align}
where $\mathbf{q}^{\mathrm{up}} = [q_1^{\mathrm{up}},\ldots,q_{2N}^{\mathrm{up}}]^T$ and $\mathbf{q}^{\mathrm{low}} = [q_1^{\mathrm{low}},\ldots,q_{2N}^{\mathrm{low}}]^T$. Thus, an iterative gradient decent method for solving~\eqref{eq_relaxed_logML_detection} can be written as
\begin{equation}
    \mathbf{x}^{(\ell)} = \mathbf{x}^{(\ell-1)} + \alpha_\ell \nabla\mathcal{P}(\mathbf{x}^{(\ell-1)}),
    \label{eq_iterative_method}
\end{equation}
where $\alpha_\ell$ is the step size of layer $\ell$. 

In order to optimize the step sizes $\{\alpha_{\ell}\}$, we use the \emph{deep unfolding} technique \cite{Hershey-Unfolding-2014} to unfold each iteration in~\eqref{eq_iterative_method} as a layer of a deep neural network. The overall structure of the proposed FBMNet is illustrated in Fig.~\ref{fig_overall_network_structure}, where there are $L$ layers and each layer takes a vector of $2K$ elements as the input and generates an output vector of the same size. The specific structure for each layer $\ell$ of FBMNet is illustrated in Fig.~\ref{fig_fewbit_structure_layer_l}. 
It can be seen that the proposed layer structure in Fig.~\ref{fig_fewbit_structure_layer_l} is different from that of conventional DNNs, since it exploits the specific structure of the reformulated ML detection problem. In particular, each layer of a conventional DNN often contains a weight matrix and a bias vector to be trained. However, due to the structure of the reformulated ML detection problem, each layer of the proposed FBMNet has two weight matrices $\mathbf{H}$ and $\mathbf{H}^T$, and two bias vectors $\mathbf{q}^{\mathrm{up}}$ and $\mathbf{q}^{\mathrm{low}}$. These weight matrices and bias vectors are defined by the channel and the received signal, respectively. Thus, they do not need to be trained. The trainable parameters in FBMNet are the $L$ step sizes $\{\alpha_1,\alpha_2,\ldots,\alpha_L\}$ and a parameter $\beta$, which acts as a scaling factor for the Sigmoid function. It should be noted that the coefficient $c\sqrt{2\rho}$ is omitted in the proposed layer structure of FBMNet since it is a constant through all layers of the network. The trainable parameters take over the role of this coefficient. Our experiments have shown that the omission of $c\sqrt{2\rho}$ not only helps improve the detection performance but also helps stabilize the convergence during training.

Since $\mathbf{H} \in \mathbb{R}^{2N\times2K}$, the learning process for each layer can be interpreted as first up-converting the signal $\mathbf{x}^{(\ell-1)}$ from dimension $2K$ to dimension $2N$ using the weight matrix $\mathbf{H}$ to obtain $\mathbf{s}^{(\ell-1)} = \mathbf{H}\mathbf{x}^{(\ell-1)}$, then applying nonlinear activation functions before down-converting the signal back to dimension $2K$ using the weight matrix $\mathbf{H}^T$. The activation function in FBMNet is the Sigmoid function, which is also widely used in conventional DNNs. Note that the use of the Sigmoid activation function in FBMNet is not arbitrary but results from the use of the approximation in~\eqref{eq_approximate_Phi_as_Sigma} and the structure of the reformulated ML detection problem. The objective function for training FBMNet is $\|\mathbf{x}^{(L)}-\mathbf{x}\|^2$, where $\mathbf{x}$ is the target signal, i.e., the transmitted signal.

\subsection{OBMNet Detector}
\label{sec_OBMNet_detector}
In this section, we propose OBMNet for one-bit massive MIMO systems. For the special case of $1$-bit ADCs, the system model can be written as
\begin{equation}
    \mathbf{y} = \operatorname{sign}(\mathbf{H}\mathbf{x} + \mathbf{z})
    \label{eq_1bit_system_model}
\end{equation}
where $\operatorname{sign}(\cdot)$ represents the $1$-bit ADC with $\operatorname{sign}(r) = +1$ if $r\geq0$ and $\operatorname{sign}(r) = -1$ if $r<0$. The ML detection problem of~\eqref{eq_1bit_system_model} is given by~\cite{choi2016near}
\begin{equation}
\hat{\mathbf{x}}_{\mathtt{ML}} = \argmax_{\bar{\mathbf{x}}\in \bar{\mathcal{M}}^{K}}\;\sum_{i=1}^{2N}\log \Phi (\sqrt{2\rho}y_i\mathbf{h}_i^T\mathbf{x}),
\label{eq_conventional_logML_1bit_detection}
\end{equation}
where $\mathbf{h}_i^T$ is the $i$-th row of the channel matrix $\mathbf{H}$. Let $\mathbf{G} = [\mathbf{g}_1, \ldots, \mathbf{g}_{2N}]^T = \operatorname{diag}(y_1,\ldots,y_{2N})\mathbf{H}$ and using the same approximation in~\eqref{eq_approximate_Phi_as_Sigma}, the ML detection problem~\eqref{eq_conventional_logML_1bit_detection} can be reformulated as
\begin{equation}
\hat{\mathbf{x}}_{\mathtt{ML}} = \argmin_{\bar{\mathbf{x}}\in \bar{\mathcal{M}}^{K}}\;\underbrace{\sum_{i=1}^{2N}\log \left(1+e^{-c\sqrt{2\rho}\mathbf{g}_i^T\mathbf{x}}\right)}_{\tilde{\mathcal{P}}_{1\mathrm{bit}}(\mathbf{x})}.
\label{eq_reformulated_1bit_ML_detection}
\end{equation}
It is interesting to note that $\log(1+e^t)$ is referred to as the SoftPlus activation function in the machine learning literature. Hence, the reformulated ML detection problem in~\eqref{eq_reformulated_1bit_ML_detection} can be interpreted as a minimization problem whose objective is a sum of SoftPlus activation functions.

The gradient of $\tilde{\mathcal{P}}_{1\mathrm{bit}}(\mathbf{x})$ is
\vspace{-0.1cm}
\begin{align}
\nabla \tilde{\mathcal{P}}_{1\mathrm{bit}}(\mathbf{x}) & = \sum_{i=1}^{2N}\frac{-c\sqrt{2\rho}\,\mathbf{g}_i}{1+e^{c\sqrt{2\rho}\,\mathbf{g}_i^T\mathbf{x}}} \nonumber\\
& = -c\sqrt{2\rho}\mathbf{G}^T \sigma \big(-c\sqrt{2\rho}\mathbf{G}\mathbf{x}\big).
\end{align}
An iterative gradient decent method for solving~\eqref{eq_reformulated_1bit_ML_detection} can also be written as
\vspace{-0.15cm}
\begin{equation}
    \mathbf{x}^{(\ell)} = \mathbf{x}^{(\ell-1)} - \alpha_\ell \nabla\mathcal{P}_{1\mathrm{bit}}(\mathbf{x}^{(\ell-1)}).
    \label{eq_1bit_iterative_method}
\end{equation}
Our proposed OBMNet approach for $1$-bit massive MIMO systems uses the same unfolding approach as FBMNet. The overall structure of OBMNet is similar to that of FBMNet as illustrated in Fig.~\ref{fig_overall_network_structure}. The specific layer structure of OBMNet given in Fig.~\ref{fig_1bit_structure_layer_l} is however simpler than the layer structure of FBMNet. In particular, each layer of OBMNet contains two adaptive weight matrices, but no bias vectors. Information about the received signal is integrated directly into the weight matrices. The number of Sigmoid functions in each layer of OBMNet is only half of that in each layer of FBMNet. The set of trainable parameters in OBMNet is similar to that of FBMNet, i.e., $L$ step sizes $\{\alpha_\ell\}$ and a scaling parameter $\beta$. The objective function for training OBMNet is $\|\tilde{\mathbf{x}}-\mathbf{x}\|^2$, where 
\begin{equation*}
    \tilde{\mathbf{x}} = \frac{\sqrt{K}}{\|\mathbf{x}^{(L)}\|}\mathbf{x}^{(L)},
\end{equation*}
and $\mathbf{x}$ is the target signal. Unlike the objective function in FBMNet, the objective function of OBMNet involves the normalization of the output of the last layer. Our simulations have shown that this normalization step significantly improves the performance of OBMNet.

\subsection{Computational Complexity Analysis}
\label{sec_complexity}
The number of real multiplications for each layer of OBMNet and FBMNet are $8KN+2N+2K$ and $8KN+4N+2K$, respectively. Therefore, the computational complexity of both OBMNet and FBMNet is of order $\mathcal{O}(KNL)$, and thus scales linearly with the number of users, the number of receive antennas, and the number of network layers.

\section{Numerical Results}
\label{sec_numerical_results}
This section presents numerical results to show the performance of the proposed OBMNet and FBMNet detectors. The channel elements are assumed to be i.i.d. and drawn from the normal distribution $\mathcal{CN}(0,1)$. The training process for both OBMNet and FBMNet is first accomplished offline. A training sample can be obtained by randomly generating a channel matrix $\mathbf{H}$, a transmitted signal $\mathbf{x}$, and a noise vector $\mathbf{z}$. The channel matrix $\mathbf{H}$ and the received signal $\mathbf{y}$ are used to define the weight matrices in OBMNet, and the weight matrices and the bias vectors in FBMNet. The transmitted signal $\mathbf{x}$ is used as the target. After the offline training process, the trained step sizes $\{\alpha_{\ell}\}$ and the trained scaling parameter $\beta$ are used for the online detection phase. Similar to DetNet for unquantized MIMO detection~\cite{Samuel2019Learning}, the proposed OBMNet and FBMNet networks do not need to be retrained for a new channel realization $\mathbf{H}$. We use TensorFlow~\cite{tensorflow2} and the Adam optimizer~\cite{kingma2014adam} with a learning rate of $10^{-2}$. The size of each training batch is set to $1000$. The input signal is set to $\mathbf{x}_0 = \mathbf{0}$.

\begin{figure}[t!]
    \centering
    \includegraphics[width=0.95\linewidth]{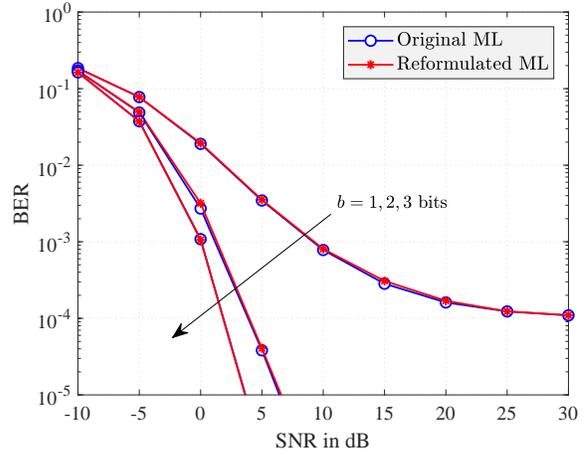}
    \caption{Performance comparison between the original ML and the reformulated ML detection problems.}
    \label{fig_ML_BERs}
\end{figure}

First, we compare the performance of the original ML detection approaches~\eqref{eq_conventional_logML_detection} and~\eqref{eq_conventional_logML_1bit_detection} with the reformulated ML detection approaches~\eqref{eq_reformulated_logML_detection} and~\eqref{eq_reformulated_1bit_ML_detection}. The BER performance is calculated via an exhaustive search for the optimal solutions and is shown in Fig.~\ref{fig_ML_BERs}. It can be clearly seen that the reformulated ML detection problems attain nearly identical performance to the original ML detection problems. It is also observed that there is a significant performance improvement as the ADC resolution increases from $1$-bit to $2$-bits.

\begin{figure}[t!]
    \centering
    \includegraphics[width=0.95\linewidth]{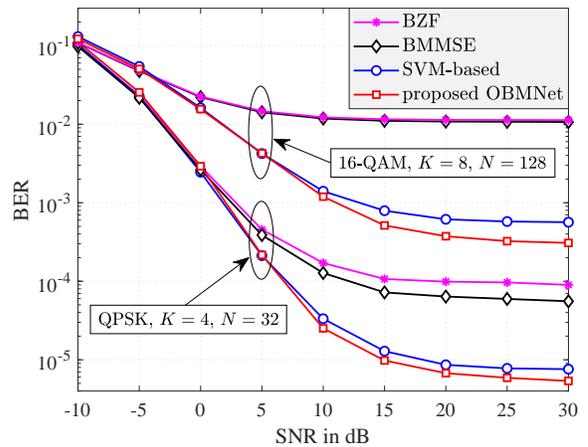}
    \caption{Performance comparison between the proposed OBMNet detector and existing methods. The number of layers $L$ in OBMNet is set to be $10$ and $15$ for the case of QPSK and $16$-QAM, respectively.}
    \label{fig_1bit_BERs}
\end{figure}
Fig.~\ref{fig_1bit_BERs} provides a performance comparison between the proposed OBMNet detector and the existing detection methods BMMSE and BZF from~\cite{nguyen2019linear}, and the SVM-based method in~\cite{Nguyen2020SVM}. The performance of OSD is comparable to that of the SVM-based approach but with a much higher computational complexity. Since the SVM-based method also outperforms other prior methods, we use it as a benchmark for OBMNet. The results in Fig.~\ref{fig_1bit_BERs} show that OBMNet and the SVM-based method outperform the Bussgang-based linear receivers. At high SNRs, the BER floor of OBMNet is slightly lower than that of the SVM-based method. The complexity of the SVM-based method is $\mathcal{O}(KN\kappa(N))$, where $\kappa(N)$ is empirically reported to be a super-linear function of $N$~\cite{joachims2006training}. Hence, OBMNet has lower complexity compared to the SVM-based method.

\begin{figure}[t!]
     \centering
     \begin{subfigure}[t]{0.49\linewidth}
         \centering
         \includegraphics[width=\linewidth]{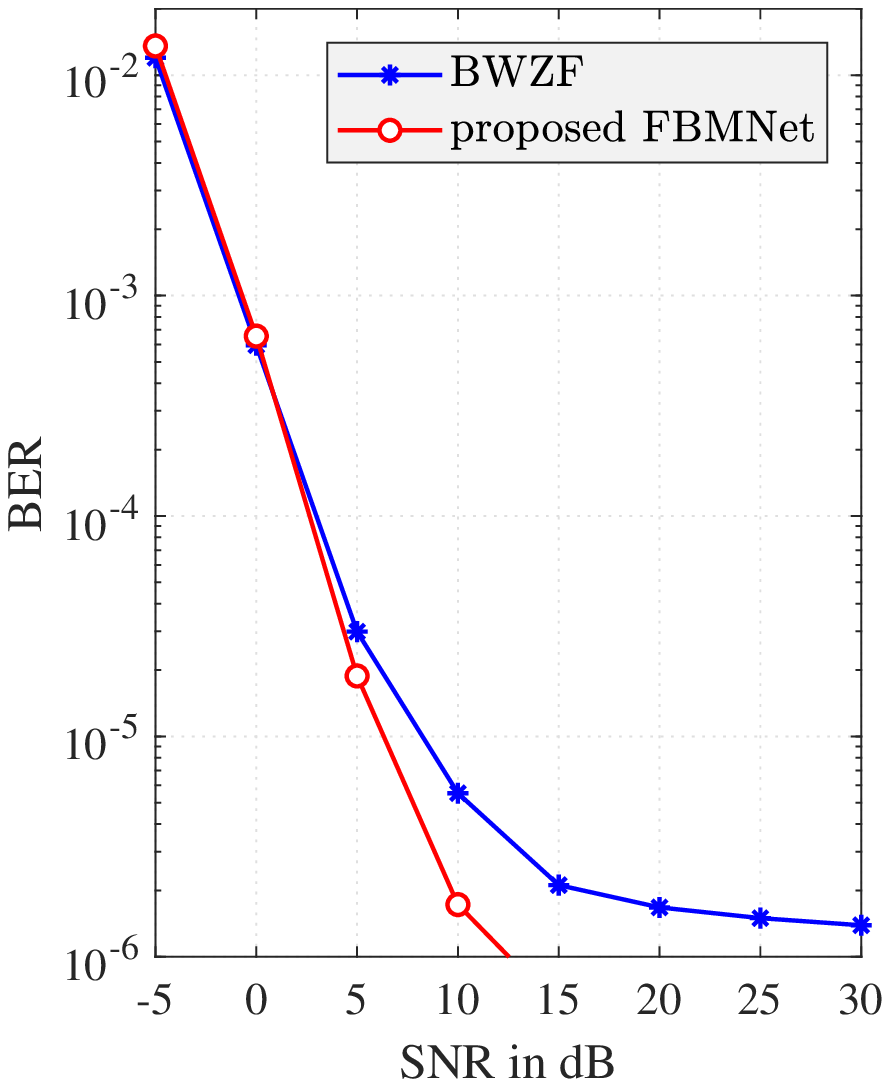}
         \caption{$b=2$ bits, $K=8$, $L=10$.}
         \label{fig_2bits_BERs_QPSK}
     \end{subfigure}~
     \begin{subfigure}[t]{0.49\linewidth}
         \centering
         \includegraphics[width=\linewidth]{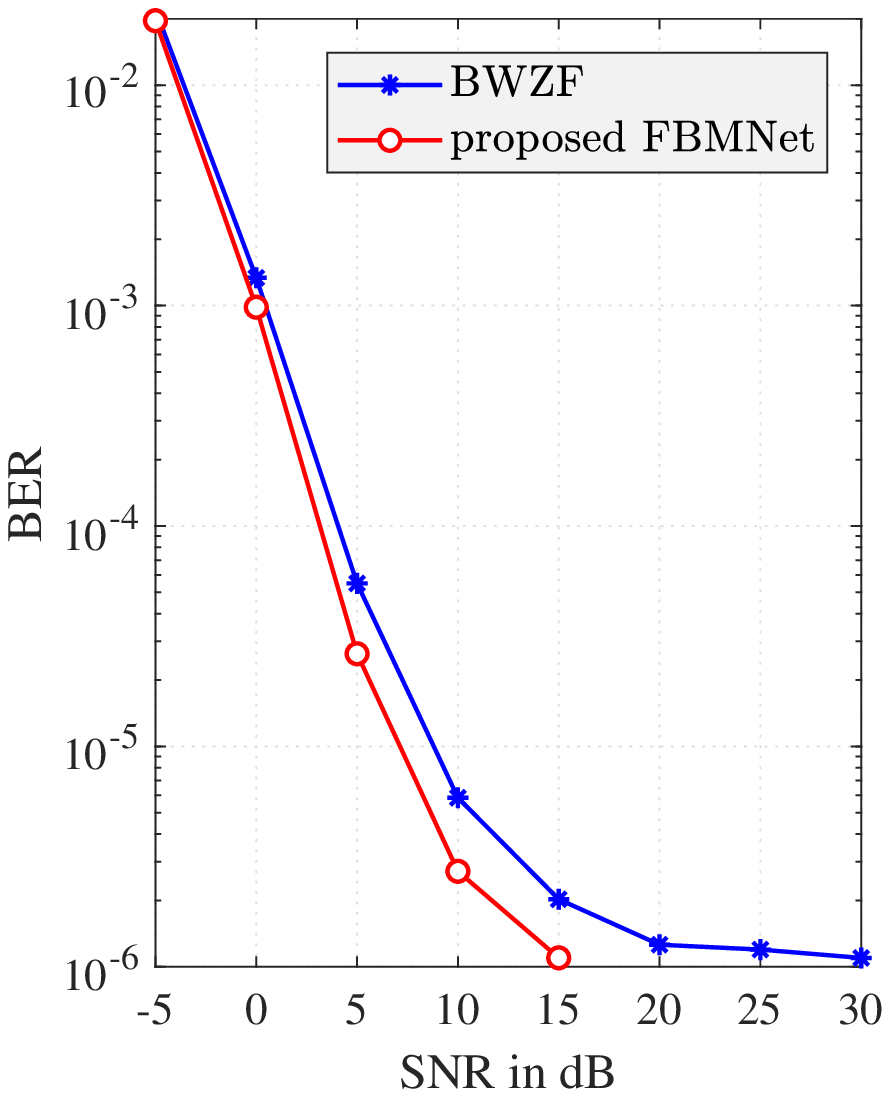}
         \caption{$b=3$ bits, $K=16$, $L=18$.}
         \label{fig_3bits_BERs_QPSK}
     \end{subfigure}
    \caption{Performance comparison between BWZF~\cite{Kolomvakis2020Quantized} and the proposed FBMNet with QPSK modulation and $N=32$.}
    \label{fig_QPSK_BERs}
\end{figure}

\begin{figure}[t!]
     \centering
     \begin{subfigure}[t]{0.49\linewidth}
         \centering
         \includegraphics[width=\linewidth]{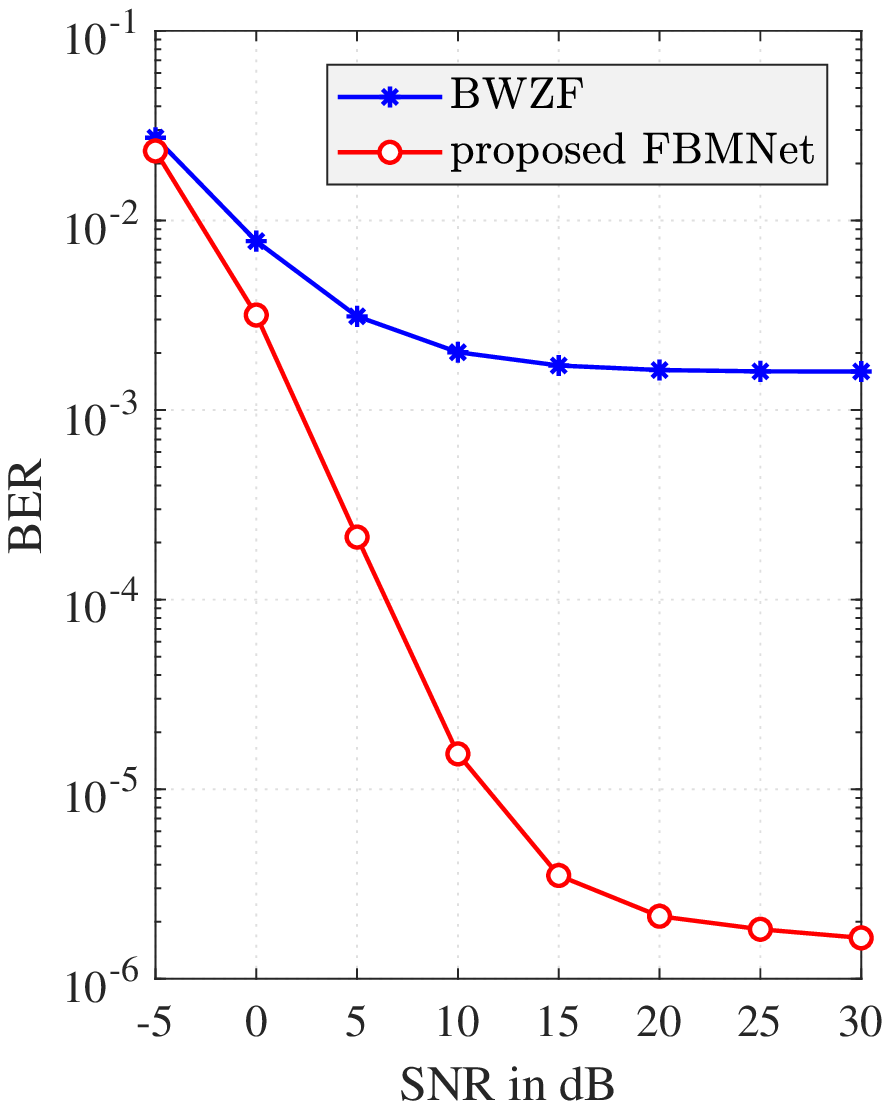}
         \caption{$b=2$ bits, $K=16$, $L=8$.}
         \label{fig_2bits_BERs_16QAM}
     \end{subfigure}~
     \begin{subfigure}[t]{0.49\linewidth}
         \centering
         \includegraphics[width=\linewidth]{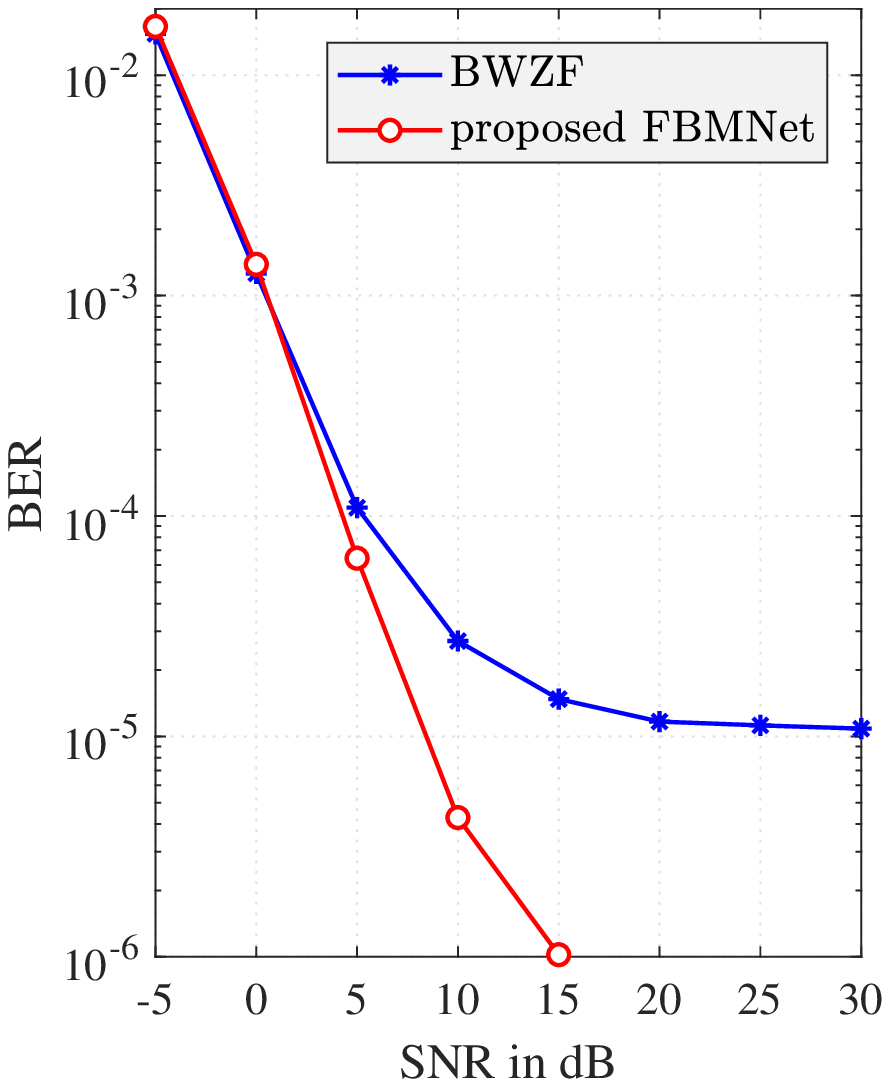}
         \caption{$b=3$ bits, $K=32$, $L=15$.}
         \label{fig_3bits_BERs_16QAM}
     \end{subfigure}
    \caption{Performance comparison between BWZF~\cite{Kolomvakis2020Quantized} and the proposed FBMNet with $16$-QAM modulation and $N=128$.}
    \label{fig_16QAM_BERs}
\end{figure}

Fig.~\ref{fig_QPSK_BERs} and Fig.~\ref{fig_16QAM_BERs} compare the proposed FBMNet detector with a recent detection method referred to as BWZF~\cite{Kolomvakis2020Quantized}. Since BWZF is reported to outperform other existing methods~\cite{Kolomvakis2020Quantized}, we use BWZF as a comparative benchmark for FBMNet. The results in Fig.~\ref{fig_QPSK_BERs} and Fig.~\ref{fig_16QAM_BERs} show that FBMNet significantly outperforms BWZF, especially with a high-dimensional constellation like $16$-QAM.

\vspace{-0.3cm}
\section{Conclusion}
\label{sec_conclusion}
In this paper, we have proposed  the novel, efficient, and low-complexity DNN-based detectors OBMNet and FBMNet for one-bit and few-bit massive MIMO systems, respectively. These proposed DNN-based detectors are model-driven and based on reformulated ML detection problems. The layered structure of OBMNet and FBMNet is simple, unique, and adaptive to the CSI and the received signals. Numerical results show that the proposed networks significantly outperform existing detection approaches.


\vspace{-0.15cm}
\bibliographystyle{IEEEtran}

\end{document}